\documentstyle[prl,aps,epsf,amsfonts,amssymb,multicol,epsfig]{revtex}

\begin{document}

\twocolumn[\hsize\textwidth\columnwidth\hsize\csname@twocolumnfalse\endcsname
 \title{Reflectivity and Microwave Absorption in Crystals with Alternating 
Intrinsic Josephson Junctions}
\author{Ch.  Helm$^{1,2}$, L.N. Bulaevskii$^{1,3}$, E.M. Chudnovsky$^3$, and
M.P. Maley$^1$ }
\address{$^1$Los Alamos National Laboratory, Los Alamos, NM 87545 \\
$^2$Institut f{\"u}r Theoretische Physik,  ETH H{\"o}nggerberg, 
8093 Z{\"u}rich, Switzerland  \\
$^3$Physics Department, CUNY Lehman College, Bronx, NY 10468-1589
}
\date{\today}
\maketitle

\begin{abstract}
We compute the frequency and magnetic field dependencies of the 
reflectivity $R(\omega)$ in layered superconductors with two alternating 
intrinsic Josephson 
junctions with different critical current densities and quasiparticle 
conductivities for the electric field polarized along the $c$-axis.
The parameter $\alpha$ 
describing the electronic compressibility of the layers and the charge 
coupling of neighboring junctions was extracted 
for the SmLa$_{1-x}$Sr$_{x}$CuO$_{4-\delta}$ superconductor
from two independent optical measurements,  the 
fit of the loss function $L(\omega)$ at zero magnetic field and 
the magnetic field dependence of the peak positions in $L(\omega)$.
 The experiments are consistent with a free electron value for 
$\alpha$ near the Josephson plasma frequencies. 
\end{abstract}
]

The Josephson plasma resonance (JPR) observed in the microwave
absorption \cite{tsui}, and in  optical reflection 
\cite{marel,shibata,kakeshita,dulic} and transmission measurements 
\cite{pim,tors}, has proven to be
a powerful method to study the properties of 
highly anisotropic layered superconductors, such as vortex phases
 \cite{tsui,kosh}. 
In particular, the spatial dispersion of the JPR, $\omega_p({\bf k},k_z)$, 
parallel (${\bf k}$) and perpendicular ($k_z$) of the superconducting layers 
reflects the inductive coupling of junctions due to 
intralayer currents and the charge coupling due to the variations of the 
electrochemical potential on the layers, respectively. 
Determining the latter is essential for understanding the coupled dynamics
of the stack of intrinsic Josephson junctions in cuprate superconductors, 
e.g. with respect to coherence in THz emission \cite{tach,sendai}. 
From a fundamental point of view, it contains unique information about 
the electronic structure of the superconducting CuO$_2$-layers, 
namely their compressibility, which is hard to obtain otherwise.
Beyond this, from the damping of the JPR  the $c$-axis 
quasiparticle (QP) conductivity at the JPR frequency can be extracted. 

The parameter $\alpha$ characterizing the $c$-axis dispersion of the JPR in 
crystals with identical junctions is difficult to observe in  bulk
microwave, transport \cite{sendai} or optical experiments, 
because mainly modes with small $k_z$ are excited,  although for grazing 
incidence the JPR peak amplitude can depend on $\alpha$ \cite{bh}.  The latter
 is irrelevant for 
the experiments in \cite{marel,shibata,kakeshita,dulic}, which are performed 
with incidence parallel to the layers, which will be studied here. 
In the following we will show, how $\alpha$ can be extracted 
unambiguously for layered superconductors with a superstructure in 
$c$-direction from two independent measurements of the loss function and the 
magnetic field dependence of the plasma frequencies. 

Recently, these experiments have been performed on 
SmLa$_{1-x}$Sr$_{x}$CuO$_{4-\delta}$, where magnetic Sm$_2$O$_2$ and 
nonmagnetic La$_{2-x}$Sr$_x$O$_{2-\delta}$ layers alternate in the barriers 
between the CuO$_2$-layers \onlinecite{shibata,kakeshita,dulic,pim}.  
In Refs. \onlinecite{shibata,kakeshita,dulic} the reflectivity $R$ and 
transmission of the electromagnetic
 wave propagating along the layers and with the electric field along the
$c$-axis (cf. Fig.~1a) were used to extract the effective dielectric 
function $\epsilon_{\rm eff}(\omega)$ with the help of the Fresnel formulas,
e.g. $R(\omega)=
[1-\sqrt{\epsilon_{\rm eff}(\omega)}]/[1+\sqrt{\epsilon_{\rm eff}(\omega)}]$.
The authors reported  two peaks 
with quite similar widths in the loss function, 
$L(\omega)=~$Im$~[-1/\epsilon_{\rm eff}(\omega)]$, and a quite large ratio 
of the peak intensities  $L(\omega_1)/L(\omega_2)$ between 10 and 20,
 see Fig.~2. 
In a two-junction model with two different plasma frequencies, but identical 
quasiparticle conductivities and no  charge coupling
 the peak amplitudes in the loss function are quite similar, see 
Fig.~2 at $\alpha=0$. This  
disagreement with the experimental data cannot be explained by 
a strong frequency dependence of the conductivities, as the widths 
of the two peaks are too similar.  
Instead, it was argued in \cite{dulic,dirk} that the $c$-axis 
coupling  plays a crucial role for the peak intensities. 
\begin{figure}
\epsfig{file=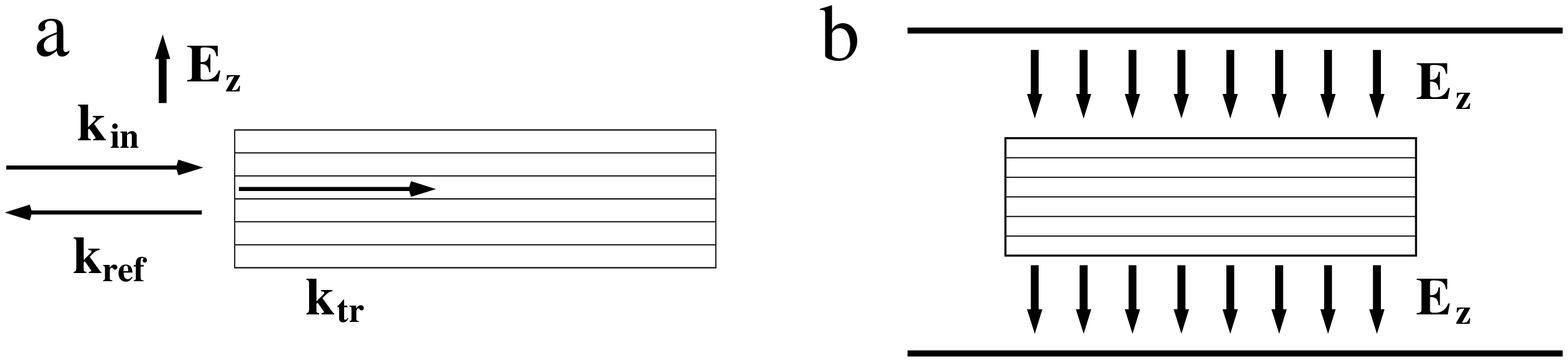,width=0.45\textwidth,clip=}
\caption{Geometry for measuring (a) the reflectivity with parallel
incidence  and (b) the microwave absorption in a cavity 
with the electric field polarized along the $z$-axis.}
\end{figure}
However, in the derivation of $\epsilon_{\rm eff}$ in Ref.~\onlinecite{dirk} 
dissipation was not introduced  in the Maxwell equations,  but arbitrarily 
in the final expression for $\epsilon_{\rm eff}$ calculated without 
dissipation.  In this Letter we correct these results accounting 
for different tunneling conductivities in the junctions in accordance with 
their different critical current densities.  We compare the theoretical 
$L(\omega)$ with the experimental data in \cite{shibata,gorshunov} and 
extract at the JPR peaks the parameter $\alpha \approx 0.4$,
 which is the free electron value. 
 Further, we show  that the microwave absorption  in the spatially 
uniform AC electric field applied along the $c$-axis (see Fig.~1b) 
is also determined by $L(\omega)$. 

Recently, Pimenov et al. \cite{pim} measured the 
dependence of the plasma frequencies 
on the magnetic field ${\bf B}\parallel c$.   
We show that the field dependence of the peak positions in $L(\omega)$ alone 
allows us to extract  $\alpha \approx 0.4$, which is an 
independent confirmation of the fit of $L(\omega)$ for $B=0$. 

We also show that the QP conductivities extracted at two different frequencies 
 do not show the $\omega$-dependence as anticipated for gapless 
$d$-wave 
pairing.  We argue that the properties of the Cooper pair tunneling via the
magnetic Sm ions may overshadow the $\omega$-dependence of the 
QP conductivity.
\begin{figure}
\epsfig{file=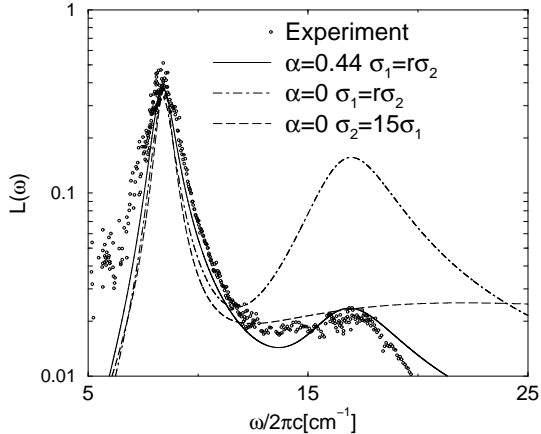,width=0.33\textwidth,angle=-90,clip=}
\caption{
Experimental data [13] together with the calculated loss function $L(\omega)$ 
(solid: $\alpha=0.44$, ${\tilde \sigma}_1 = r 
{\tilde \sigma_2}=0.12$, $r=0.38$, $\omega_{0,1}/ c=7.2~{\rm cm}^{-1}$, 
$\epsilon_{0}=18$; 
dot-dashed: $\alpha=0$, ${\tilde \sigma}_1 = r 
{\tilde \sigma_2}=0.12$, $r=0.24$, $\omega_{0,1}/c=8.3~{\rm cm}^{-1}$, 
$\epsilon_{0}=13$; dashed: $\alpha=0$, ${\tilde \sigma}_2 = 15 
{\tilde \sigma_1}=0.2$, $r=0.24$, $\omega_{0,1}/ c=8~{\rm cm}^{-1}$, 
$\epsilon_{0}=15$). 
\label{losspic}
}
\end{figure}
We consider the crystal with two alternating Josephson junctions
characterized by different critical current densities, $J_{l}$, and the 
$c$-axis tunneling conductivities, $\sigma_l$, due to 
different tunneling matrix elements.  
We assume that all other parameters of the
junctions are identical, as their distinction would lead to negligible 
corrections in the following.  Without the charging effect the 
$c$-axis bare plasma frequencies $\omega_{0,l}$ are related to $J_{l}$ as
$ \omega_{0,l}^2=8\pi^2csJ_{l}/\epsilon_{0}\Phi_0$, 
where $\Phi_0$ is the flux quantum, $\epsilon_{0}$ is the high
frequency $c$-axis dielectric constant and $s$ is the interlayer distance.  
We neglect nonequilibrium effects \cite{sendai,art}
in the distribution function of the quasiparticles assuming that the 
plasma frequencies are well below the charge imbalance and energy 
relaxation rates. 

To find the reflectivity $R(\omega)$ in parallel incidence  and the 
microwave absorption, we use the 
Maxwell equations inside the crystals accounting for supercurrents inside the
2D layers at $z=ms$ and interlayer Josephson and quasiparticle currents 
determined by the difference of the electrochemical potentials in 
neighboring layers: 
\begin{eqnarray}
&&c\frac{\partial B_y}{\partial z}=
i\epsilon_{a0}\omega\left[E_x-\frac{\omega_{a0}^2}
{\omega^2}\sum_{m=1}^{N}E_xs \delta (z-ms)\right], \label{first} \\
&&\frac{\partial E_x}{\partial z}-ik_xE_z=i \frac{\omega}{c} B_y, 
\ \ E_{z,m,m+1}= \int_{ms}^{(m+1)s}   E_z\frac{dz}{s}, \\
&&ck_xB_y=-\omega\epsilon_0\left[
E_z-\sum_{m=1}^NA_mf_{m,m+1}(z)\right], \label{e} \\ 
&& \frac{{\tilde \omega}_l^2 es}{\omega_{0,l}^2} A_m= V_{m,m+1} 
=esE_{z,m,m+1}+\mu_{m+1}-\mu_{m}. \label{last}
\end{eqnarray}
Here  $\mu_m$ is the chemical potential in the layer $m$, 
$V_{m,m+1}$ is the difference of the electrochemical potentials, 
$\omega_{a0}=c/\lambda_{ab}\sqrt{\epsilon_{a0}}$ is the in-plane plasma 
frequency and $\epsilon_{a0}$ is the high frequency in-plane dielectric 
constant. The function $f$ is defined as
$f_{m,m+1}(z)=1$ at $ms<z<(m+1)s$ and zero outside this interval.
To obtain Eq.~(\ref{e}) for small amplitude
oscillations we expressed the supercurrent density
$J_{m,m+1}^{(s)}=J_l\sin\varphi_{m,m+1}\approx J_l\varphi_{m,m+1}$ 
via the phase difference $\varphi_{m,m+1}=2iV_{m,m+1}/\hbar\omega$.
Further, 
$\tilde{\omega}_l^2=\omega^2(1-i4\pi\sigma_l \omega / \omega_{0,l}^2
\epsilon_{0})^{-1}$ takes into account the
dissipation due to QP tunneling currents, 
$J_{m,m+1}^{(qp)}=\sigma_lV_{m,m+1}/es$, which are depend on the different 
conductivities $\sigma_l$  in the junctions $l=1,2$ and the difference 
$V_{m,m+1}$ of the {\it electrochemical} potentials. 
We express the difference of the chemical potentials $\mu_m$
 via the difference of the 2D charge densities, ${\rho}_m$, as 
$\mu_m-\mu_{m+1}=(4\pi  s\alpha/\epsilon_0)({\rho}_m-{\rho}_{m+1})$, 
where the 
parameter $\alpha=(\epsilon_0/4\pi e s)(\partial \mu/\partial {\rho})$ 
characterizes the interlayer coupling.  In the model of 2D free electrons 
we find $\partial \mu/\partial {\rho}=\pi\hbar^2/(e m^*)$. For an effective 
mass  $m^*\approx 1-2 m_e$, as expected from ARPES \cite{norman},
 $s=6.3$ ${\rm \AA}$  
and $\epsilon_0=20$ we  estimate the order of $\alpha$ as 
$\approx 0.2-0.4$.

The solution of the Eqs.~(\ref{first})-(\ref{last}) is 
\begin{eqnarray}
&&B_y(z)=\frac{\epsilon_{0}\omega}{ck_x}[c_m\exp(igz)+d_m\exp(-igz)]
-a\frac{ck_x}{\omega} A_m, \nonumber \\
&&E_x (z)=\frac{\epsilon_{0}\omega}{(\epsilon_{a0}a)^{1/2}ck_x}
[c_m\exp(igz)-d_m\exp(-igz)], \nonumber \\
&&E_z(z)=-[c_m\exp(igz)+d_m\exp(-igz)] +aA_m,
\end{eqnarray}
where now $0\leq z\leq s$ in each junction, 
$g^2=\omega^2 \epsilon_{a0}/(c^2 a)$ and $a^{-1}=1-c^2k_x^2/\epsilon_{0}
\omega^2$.  
The Maxwell boundary conditions for $B_y$ and $E_x$ across the layers 
lead to a set of equations for $c_m$, $d_m$ and $A_m$, which we solve 
using the small parameters $b=gs/2 \sim s/\lambda_c\ll 1$ and 
$\beta=s^2/2\lambda_{ab}^2 a \ll 1$. 
In order to find the reflectivity for parallel incidence, cf. Fig.~1a, 
we consider the case $sk_z\ll b \lesssim \beta$, which is fulfilled for 
incident angles $\theta\ll 1$, and obtain 
the dispersion relation $k_x(\omega)$ for waves 
propagating inside the crystal (for details see \cite{longpaper}):
\begin{eqnarray}
&&\frac{c^2k_x^2}{\omega^2\epsilon_{0}}=
\frac{\epsilon_{\rm eff}(w)}{\epsilon_{0}}=
\frac{r(w-v_1)(w-v_2)+iS}{r w^2-(1+r)(2\alpha+1/2)w+iS_1}, 
\label{r}\\
&&v_{1,2}=(1+r)(1+2\alpha)(1\mp \sqrt{1-p})/2r,  
\label{v} \\
&&p=\frac{4r(1+4\alpha)}{(1+r)^2(1+2\alpha)^2},  
\label{p} \\
&&S_1=w^{3/2} r (2 \alpha + 1/2) ({\tilde \sigma}_1 + 
{\tilde \sigma}_2) , \\
&&S=w^{1/2} [(2\alpha+1) r w (\tilde{\sigma}_1+{\tilde \sigma}_2)-
(1+4\alpha)(\tilde{\sigma}_1+\tilde{\sigma}_2r)], \nonumber
\end{eqnarray}
where $v_l=\omega^2_l/\omega^2_{0,1}$, $\omega_l$ are the JPR frequencies, 
$w=\omega^2/\omega_{0,1}^2$, 
$\tilde{\sigma}_l=4\pi\sigma_l/\epsilon_0\omega_{0,1}$ and $r=
\omega_{0,1}^2/\omega_{0,2}^2<1$.  
\begin{figure}
\epsfig{file=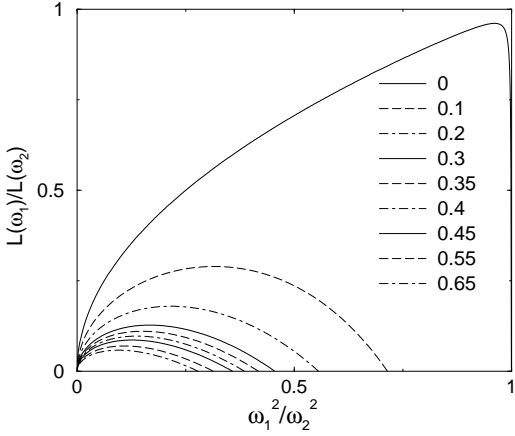,width=0.33\textwidth,angle=-90,clip=}
\caption{
\label{ratiopic}        
Ratio of the peak amplitudes $L(\omega_1)/L(\omega_2)$ in the loss function 
 depending on the squared ratio 
$\omega_1^2/ \omega_2^2$ of the JPR frequencies for different 
$\alpha=0, \dots, 0.65$ (top to bottom plot) at
${\tilde \sigma}_1 = r {\tilde \sigma}_2$.
 }
\end{figure}
\begin{figure}
\epsfig{file=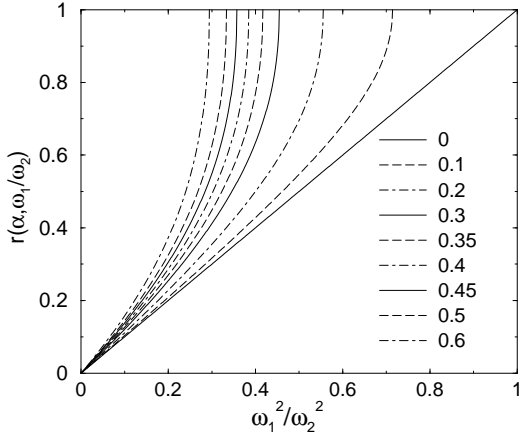,width=0.33\textwidth,angle=-90,clip=}
\caption{
\label{deltaalpha}      
Ratio $r = \omega_{0,1}^2/\omega_{0,2}^2$ of the squared
 bare plasma frequencies 
depending of the squared 
ratio $\omega_{1}^2/\omega_{2}^2$ of the JPR resonances in 
$L(\omega)$ for different $\alpha=0,..,0.6$ (from below). 
 }
\end{figure}
The reflectivity coefficient is given by the usual Fresnel expression 
$R=(1-ck_x/\omega)/(1+ck_x/\omega)$.  Note that for nonzero
$\beta$ we obtain in principle two propagating modes inside the crystal,
 but the second 
mode gives a negligible contribution near the JPR peaks at $\beta\ll 1$, see
\cite{longpaper}.  The loss function is 
\begin{eqnarray}
L(w)&=&~{\rm Im}~[-1/\epsilon_{\rm eff}(\omega)]=
(w^{3/2}/ 2 \epsilon_{0}) \times \label{ab} \\
&&\frac{{\tilde \sigma}_1(wr-4\alpha-1)^2+
{\tilde \sigma}_2r^2(w-4\alpha-1)^2}{[r(w-v_1)(w-v_2)]^2+S^2}
\nonumber 
\end{eqnarray}
and shows resonances at the two transverse plasma bands, while 
the peak in Im$[\epsilon_{{\rm eff}}(\omega)]$ is at 
$v_T=(1+1/r)(2\alpha+1/2)$.

In $s$-wave superconductors with nonmagnetic ions in the barrier between 
the layers both the QP 
conductivities $\sigma_l$ and the critical current densities $J_{l}$ vary 
with $l$ only due to their proportionality to the squared matrix element for
the interlayer tunneling in 
agreement with the doping dependence 
$\sigma_c(x)\propto \omega_p^2(x)$ found for 
La$_{2-x}$Sr$_x$CuO$_4$ \cite{uch}. 
We take the same relation, $\sigma_1 (\omega_1)/\sigma_2 (\omega_2)=r$ for 
SmLa$_{1-x}$Sr$_{x}$CuO$_{4-\delta}$, 
 ignoring  the frequency dependence of $\sigma_l$ due to the 
$d$-wave pairing \cite{hirsch}, which decreases $\sigma_1 
(\omega_1)/\sigma_2 (\omega_2)$ below $r$. On the other hand, we also 
neglect that due to the  magnetism of the Sm ions  the 
critical current density $J_1 \propto \omega_{0,1}^2$ is suppressed, 
while $\sigma_l$ is not affected, which suggests $\sigma_1/\sigma_2 >r$ 
without $d$-wave pairing \cite{bula}. 
The fact that our assumption $\sigma_1 = r \sigma_2$ turns out to 
be consistent with the experimental data suggests that both effects compensate
each other.

In Fig.~\ref{ratiopic} we present the dependence of the ratio of the peak 
amplitudes in the loss function $L(\omega)$, i.e. $L(\omega_1)/L(\omega_2)$, 
vs. the ratio $\omega_1^2/\omega_2^2$ 
for different values of $\alpha$. Hence, this figure allows us to 
obtain the parameter $\alpha$ from the positions $\omega_{1,2}$
of the JPR peaks and their ratio of amplitudes. 
Fig.~\ref{deltaalpha} allows us to find the ratio 
of the squared bare frequencies, $r=\omega_{0,1}^2/\omega_{0,2}^2$ for 
a ratio $\omega_1/\omega_2$ when the parameter 
$\alpha$ is obtained from Fig.~\ref{ratiopic}. 
The comparison of $L(\omega)$
 in Fig.~\ref{losspic} with the experimental data 
from Ref.~\onlinecite{gorshunov} gives the best fit for 
${\tilde \sigma}_1 = r {\tilde \sigma}_2=0.12$ and $\epsilon_{0}=18$, 
 $\alpha=0.44$, which is of the same order as the theoretical 
 estimate for free electrons, cf. Tab. I for a fit of 
\cite{shibata,dulic,pim,gorshunov}. In the case $\alpha=0$ a high ratio 
${\tilde \sigma}_2 / {\tilde \sigma}_1 \approx 15$ of the conductivities is 
necessary to reproduce the ratio of the amplitudes, which fails to 
describe the shape of the resonance at $\omega_2$ correctly.

The loss function $L(\omega)$ also determines the microwave absorption of 
a crystal in  a 
capacitor, which induces a uniform ($sk_z\rightarrow 0$, $k_x=0$) AC 
electric field $2{\cal E}\cos(\omega t)$ above and 
below the crystal, see Fig.~1b.  The 
crystal excitations are longitudinal in this case,
i.e.  $E_x=B_y=0$ 
in Eqs.~(\ref{first})-(\ref{last}). 
These equations together with the Poisson equations near the top and bottom 
layers determine for large $N$  the microwave absorption as 
\begin{equation}
{\cal P}(\omega)=N^{-1}\sum_{m=1}^{N} \sigma_l|V_{m,m+1}/es|^2 
=\frac{\omega}{4\pi}L(\omega) {\cal E}^2 .
\end{equation} 

Next we consider the dependence of the JPR peaks in $L(\omega)$ on the 
$c$-axis magnetic field $B$, which allows one to estimate 
the parameter $\alpha$, without relying on the absolute amplitude of the 
spectra.  
For Josephson junctions $B$
 suppresses the critical current densities and the
bare plasma frequencies, $\omega_{0,l}(B)$, and also broadens the 
resonance peaks due to the formation of
pancake vortices randomly misaligned along the $c$-axis 
\cite{kosh}.
From Eq.~(\ref{v}) we see that 
the ratio of the resonance frequencies, 
\begin{equation}
\omega_1^2/\omega_2^2=(1-\sqrt{1-p})/(1+\sqrt{1-p}), 
\label{omegarat}
\end{equation}
depends only on the quantity $p$ given by Eq.~(\ref{p}), which varies with 
$B$ only via the function $r(B)=\omega^2_{0,1}(B)/
\omega^2_{0,2}(B)$. 
When the functional form of $\omega_{0,l} (B)$ is known theoretically and 
$p(B)$ is obtained from the measurement of $\omega_1/\omega_2$ via 
Eq.~(\ref{omegarat}), one can fit unknown parameters in 
$\omega_{0,l} 
(B)$ and obtain $\alpha$. It is pointed out that extracting $\alpha$ 
from $L(\omega,B)$  by fitting all parameters without further theoretical 
input is difficult due to the dependence of the effective $c$-axis 
conductivity on $B$ \cite{kosh}. 

The dependence $\omega_{0,l}(B)$ in the decoupled 
vortex liquid \cite{kosh} is known 
for crystals with equivalent junctions and can be used for alternating 
junctions as well, provided that the vortices in different layers are 
decoupled. This is suggested by  $\hbar \omega_{0,l} \ll k T_c$ in 
SmLa$_{0.8}$Sr$_{0.2}$CuO$_{4-\delta}$ (cf. Tab. I) indicating Josephson 
coupling of the layers. Thus, for 
the vortex liquid phase with decoupled pancakes
in high magnetic fields $B\gg B_J=\Phi_0 \lambda_{ab}^2/ (\lambda_c s)^2 
\sim 1$ T the dependence of the 
plasma frequencies on $B$ is  \cite{kosh}
\begin{equation}
\omega_{0,l}^2(B)\approx\omega_{0,l}^4(0)\frac{\epsilon_{0}\Phi_0^3}{32B\pi^3
c^2sT}.
\end{equation}   
This gives $r(B\gg B_J)=r^2(B=0)$.  Hence, we find 
\begin{equation}
p(B\gg B_J)=\frac{4r^2(0)(1+4\alpha)}{[1+r^2(0)]^2(1+2\alpha)^2},
\label{ddd}
\end{equation}
while for $p(0)$ we use Eq.~(\ref{p}) with $r=r(0)$.

Consequently, the parameters $\alpha$ and $r(0)$ can be found analytically 
without fitting
from $p(0)$ and $p(B\gg B_J)$ obtained from JPR frequency measurements
 using the data of Ref. \cite{pim}.  The 
dependence $\omega_1^2,\omega_2^2\propto 1/B$ at fields above
 $B_J \approx 1$ T
 shows that in this field range the decoupled pancake liquid 
is present. We use the reported field dependence of the peak frequency in 
Im$[\epsilon_{{\rm eff}}(\omega)]$ for $\omega_2(B)$, as
the authors noted that they are quite close and the 
difference will finally turn out to be $\approx 10$ \%. Then we obtain 
(for $T \ll T_c$) 
$\omega_1^2/\omega_2^2\approx 0.31$ at $B=0$ and 0.21 at $B>B_J$  or
$p(0)\approx 0.73$ and $p(B\gg B_J)\approx 0.57$
respectively.   From Eqs.~(\ref{p}) and (\ref{ddd}) we obtain 
$r(0)\approx 0.55$, $\alpha\approx 0.4$ and the bare frequencies are 
$\omega_{0,1}/ c=6.6$ cm$^{-1}$ and $\omega_{0,2}/ c=8.9$ cm$^{-1}$,
cf. Tab. I. 
This estimate gives the parameters $\alpha$ and $r(0)$ 
similar to those obtained above from the 
fit of $L(\omega)$ at $B=0$  and confirms independently the relevance of the
$c$-axis coupling $\alpha$  and our assumption 
${\tilde \sigma}_1 \approx r {\tilde \sigma}_2$. 
\begin{table}
\begin{tabular}{c|c|c|c|c|c|c}
$x$&$T_c$&$\omega_{0,1}/ c$
&$r$&$\sigma_{1}$[$(\Omega$m)$^{-1}$]&$\alpha$&Ref.\\ \hline
0.15&23K&10.9{\rm cm}$^{-1}$&0.42&10&0.36&[3]\\ 
0.15&30K&7.2{\rm cm}$^{-1}$&0.38&4.3&0.44&[13]\\
0.2&17K&6.6{\rm cm}$^{-1}$&0.55&3.9&0.40&[5,6] 
\end{tabular}
\caption{Extracted parameters for SmLa$_{1-x}$Sr$_{x}$CuO$_{4-\delta}$  }
\end{table} 
In conclusion, we calculated the effective dielectric function 
$\epsilon_{\rm eff}$, Eq.~(\ref{r}), for  alternating junctions with charge 
coupling $\alpha$ and frequency independent QP conductivities 
$\sigma_l\propto \omega_{0,l}^2$. This allows
to describe satisfactory the optical properties of the 
SmLa$_{0.8}$Sr$_{0.2}$CuO$_{4-\delta}$ superconductor near the plasma 
frequencies. 
The parameter $\alpha\approx 0.4$ was extracted independently
from the magnetic field dependence of the 
positions of the JPR peaks $\omega_{1,2}(B)$ and from the 
fit of the loss function $L(\omega)$ at $B=0$. Its value is expected to be 
universal in the cuprates and  corresponds to the
 free electron value of the electronic compressibility, which can   
differ from the renormalized one particle density of states both for Fermi 
and non Fermi liquids  \cite{nonfermi}. 
We also show that the extracted  conductivities $\sigma_l$
differ according to the different tunneling matrix elements, 
$\sigma_1 / \sigma_2 \approx \omega_{0,1}^2 / \omega_{0,2}^2 $. This relation 
is anticipated for tunneling between $s$-wave superconductors via nonmagnetic
ions.  Its fulfillment in SmLa$_{0.8}$Sr$_{0.2}$CuO$_{4-\delta}$ 
might indicate that the  gapless $d$-wave pairing in the layers 
reduces $\sigma_1(\omega_1)/\sigma_2(\omega_2)$ in a similar way as 
the magnetism of the Sm ions decreases  $\omega_{0,1}^2/\omega_{0,2}^2$. 
This could be studied further by measuring the gap in the  QP spectrum in
the $I$-$V$ curve as in Ref.~\onlinecite{hirsch}. 

The authors thank M. Graf and W. Zwerger for useful discussions. 
The work was supported by the U.S.
DOE (in Lehman College through  Grant No. DE-FG02-93ER45487) 
and the NRCC of the Swiss NSF.

\end{document}